\documentclass[12pt,superscriptaddress,showpacs,showkeys,twocolumn,prb,aps,reprint,a4paper,floatfix]{revtex4-1}
\usepackage{graphicx}
\usepackage{amssymb,amsmath}
\newcommand*\diff{\mathop{}\!\mathrm{d}}
\newcommand*\Diff[1]{\mathop{}\!\mathrm{d^#1}}
\newcommand*\rme{\mathop{}\!\mathrm{e}}
\newcommand*\rmi{\mathop{}\!\mathrm{i}}
\newcommand*\crit{\mathop{}\!\mathrm{c}}
\newcommand*\Sl{\mathop{}\!\mathrm{S}}
\newcommand*\Sp{\mathop{}\!\mathrm{S'}}
\newcommand*\Spp{\mathop{}\!\mathrm{S''}}
\newcommand*\nn{\mathop{}\!\mathrm{n}}
\usepackage{bm}
\usepackage{physics}
\usepackage{enumerate}
\usepackage{natbib}
\begin{document}
\title[Modelling proximity effects in TESs]{Modelling proximity effects in Transition Edge Sensors to investigate the influence of lateral metal structures}

\author{R. C. Harwin}
\email{rch66@cam.ac.uk}
\author{D. J. Goldie}
\author{S. Withington}
\affiliation{Cavendish Laboratory, JJ Thomson Avenue, Cambridge CB3 OHE, United Kingdom.}

\begin{abstract}
The bilayers of Transition Edge Sensors (TESs) are often modified with additional normal-metal features such as bars or dots. Previous device measurements suggest that these features improve performance, reducing electrical noise and altering response times. However, there is currently no numerical model to predict and quantify these effects. Here we extend existing techniques based on Usadel's equations to describe TESs with normal-metal features. We show their influence on the principal TES characteristics, such as the small-signal electrothermal parameters $\alpha$ and $\beta$ and the superconducting transition temperature $T_{\crit}$. Additionally, we examine the effects of an applied magnetic field on the device performance. Our model predicts a decrease in $T_{\crit}$, $\alpha$ and $\beta$ as the number of lateral metal structures is increased. We also obtain a relationship between the length $L$ of a TES and its critical temperature, $T_{\crit} \propto L^{-0.7}$ for a bilayer with normal-metal bars. We predict a periodic magnetic flux dependence of $\alpha, \beta$ and $I_{\crit}$. Our results demonstrate good agreement with published experimental data, which also show the reduction of $\alpha$, $\beta$ and $T_{\crit}$ with increasing number of bars. The observed Fraunhofer dependence of critical current on magnetic flux is also anticipated by our model. The success of this model in predicting the effects of additional structures suggests that in the future numerical methods can be used to better inform the design of TESs, prior to device processing.
\end{abstract}

\pacs{74.45.+c, 74.50.+r, 74.78.Fk}
\keywords{Transition Edge Sensors, Long Range Proximity Effect, S-S'-S Junctions, Usadel equations}

\maketitle

\section{Introduction}
Transition Edge Sensors (TESs) are used as ultra-sensitive detectors for astronomical observations across the whole of the electromagnetic spectrum. \cite{Irwin2005,Hoover2009} They are operated at low temperatures ($T\sim 1\,\,{\rm K}$ or below), and give unparalleled performance in terms of energy resolution when used as microcalorimeters, or noise equivalent powers when used as bolometers. Various groups worldwide have achieved performances approaching theoretical limits, but understanding the factors that determine the achieved performance remains a goal. TESs are currently used in ground-based telescopes such as EBEX,\cite{Reichborn-Kjennerud2010} SPTpol,\cite{Yefremenko2013} Keck/SPIDER,\cite{Orlando2010} and GISMO, \cite{Staguhn2006} and will be used, or are proposed for use, in a number of upcoming space missions. \cite{Roelfsema2014,Gottardi2016a}

In its simplest form a TES consists of a superconducting thin-film (denoted here by $\Sp$), typically $\sim 100\,\,{\textrm{nm }}$ thick, operated very close to its superconducting transition temperature $T_{\crit,\Sp}$. This film has length $L$ and width $W$ satisfying $L{,}\, W \gg\xi$, the superconducting coherence length. The TES is coupled electrically to an external readout circuit by superconducting leads (denoted S), of higher transition temperature $T_{\crit,\Sl}$, so that Andreev reflection minimizes the electronic thermal conduction out of the superconducting film. The presence of the higher-$T_{\crit}$ leads influences the operating conditions of the TES due to the superconducting proximity effect, in which Cooper pairs formed in the leads diffuse into the lower-$T_{\crit}$ superconducting film, inducing or enhancing superconductivity.

In many TESs the superconducting film is modified, perhaps empirically, by adding features such as dots or lateral metal bars, which often only extend partially across the TES. These modifications reduce excess electrical noise\cite{Staguhn2006} and change aspects of performance such as response times.\cite{Ullom2004a,Goldie2009,Chang2012,Swetz2012a,Smith2013,Yefremenko2013} In microcalorimeters normal-metal or semi-metal structures act as, or thermally couple to, photon absorbers.\cite{Smith2014} Changes in the TES response time indicate a change in the small-signal electrothermal parameter $\alpha=(\partial \ln R(T,I) /\partial \ln T)_I$, which characterizes the sharpness of the superconducting-normal resistive transition (here $R$ is the TES resistance and $I$ the bias current). Large values of $\alpha$ optimize the current-to-power sensitivity of a TES operated as a bolometer and reduce its response time, whilst at the same time reducing its in-band Johnson noise. Changes in $R(T,I)$ alter the more easily measurable resistance-current sensitivity $\beta=(\partial \ln R/\partial \ln I)_T$.\cite{Irwin2005} Large values of $\beta$ increase the response time and also reduce the electrical stability of the TES when biased optimally at low bias voltage $V_b$.\cite{Irwin2005} Understanding the dependence of critical current $I_{\crit}$ on magnetic field $B$ as a function of TES geometry is also important in achieving ultimate power or energy sensitivity and insensitivity to environmental magnetic fields. The current to incident-power sensitivity of a bolometer is $\diff I/\diff P\approx -1/V_b$, to first order. Assuming $I=k I_{\crit}$, where the constant of proportionality $k$ is less than unity, $k \lesssim 1 $, (as in a simple two-fluid model\cite{Irwin2005}), the sensitivity of apparent detected power to field is then $\diff P/ \diff B = -k V_b\, \diff I_{\crit}/\diff B$.

From a design perspective, the \textit{effective} $T_{\crit}$ of the TES, that we define for a fixed current as the temperature at which the resistance of the actual structure equals half its normal-state value $R_\mathrm{n}$, should be engineered to match experimental requirements determined by bath temperature, power handling, power sensitivity and electrical impedance. In practice, there are few elemental superconductors with the desired characteristics such as low $T_{\crit}$, mechanical stability, ease of deposition and patterning, and resistance to environmental influences such as oxidation. Superconducting bilayers, composed of either a normal metal and a superconductor or two dissimilar superconductors, offer the flexibility to adjust $T_{\crit}$ of the superconducting film, again using the superconducting proximity effect.\cite{Martinis2000} Mo/Au, Ti/Au and Ti/Nb have all been used as bilayer combinations in this context. The use of these bilayers means that a TES can consist of a superconducting proximity structure in which the proximity effect acts in all spatial directions.
 Any additional bars or dots, denoted $\Spp$, are often formed for practical reasons (such as available thin film deposition capability) by thickening the lower-$T_{\crit}$ material of the bilayer in selected areas.

The experimental work of Sadleir {\textit{et al.}}\cite{Sadleir2010}, which strongly suggests the existence of a long-range proximity effect in MoAu TESs at low $T/ T_{\crit,\Sl}$, included the observation of a Fraunhofer-like relationship between the flux $\Phi$ and the critical current $I_{\crit}(\Phi)$, where $\Phi=BLW$ with $B$ the applied magnetic field. A similar dependence was found by Smith {\textit{et al.}}\cite{Smith2014} for MoAu TESs with partial bars and Hijmering {\textit{et al.}}\cite{Hijmering2014} for bare TiAu TESs. The measurements reported in the latter also indicate a more complicated field-dependence in MoAu TESs in large fields $\Phi / \Phi_0 \sim 50 $, where $\Phi_0=h/2e$ is the flux quantum, possibly due to the additional partial lateral metal bars.

Our brief synopsis of experimental observations is as follows:
\begin{enumerate}[i.]
\item The effective $T_{\crit}$ scales approximately as $(1/L^2)$.\cite{Sadleir2010}
\item The width of the transition  shows a similar dependence on $L$, suggesting that $\alpha$ is unchanged.
\item $\alpha$ is reduced by the addition of lower $T_{\crit}$, $\Spp$ bars (and likewise increased by higher-$T_{\crit}$ bars).\cite{Wang2015,Chang2012}
\item $\alpha$ and $\beta$ show oscillations when a magnetic field is applied. The maximum of these oscillations reduces with increasing field, so both parameters are maximised in the zero field case. \cite{Hijmering2013,Smith2013}
\item The ratio $\alpha/\beta$ is typically in the range $10 - 200$.\cite{Goldie2008,Smith2013,Lindeman2006,Rostem2010}
\item The measured $I_{\crit}(\Phi)$ shows a mostly Fraunhofer-like dependence, although the dependence is not exact.\cite{Smith2013,Gottardi2014,Sadleir2011,Hijmering2014} Some measurements on TESs with partial bars show a more  complicated field dependence.\cite{Hijmering2014}
\end{enumerate}

A general method for describing the spatial proximity effect in materials with short electronic mean-free-paths is provided by the diffusive Usadel equations.\cite{Usadel1970} These equations have been used widely to describe  S$\Sp$S  structures \textit{near} $T_{\crit,\Sl}$, including calculation of the critical current dependence on magnetic flux $I_{\crit} (\Phi)$.\cite{Likharev1979,Golubov2004,Heida1998,Vasenko2008,Chiodi2012} Cuevas and Bergeret\cite{Cuevas2007} solved the Usadel equations at low temperatures $T/T_{\crit} \sim 0.01$ in two dimensions for short wires $L/\xi=2$ with varying widths $0.5\le W / \xi \le 50$, finding a Fraunhofer-like dependence of $I_{\crit}(\Phi)$ for wide junctions $W / \xi \sim 50$. Kozorezov {\textit{et al.}} extended the modelling to long one-dimensional superconducting structures $L/\xi_{\Sp} \gg 1$ in simple S$\Sp$S geometries at low temperatures $T/T_{\crit,\Sl} \ll 1$, and showed how to describe a TES in the context of the resistively shunted Josephson junction (RSJ) model.\cite{Kozorezov2011,Kozorezov2011a} This latter work we refer to here as the ``1-D model".

Here we extend the 1-D model to explore numerically the effects of additional structures, such as bars, calculating how these change the observed $T_{\crit}$, the electrothermal parameters $\alpha$ and $\beta$, and the current-field dependence $I_c(\Phi)$. We calculate the effect of magnetic field on $\alpha$ and $\beta$. We describe the generic effects of changes in the number of bars or the length $L$ on the electrothermal parameters.
To enable this work we also describe how to account for changes in {\textit{thickness}} arising from the structures, taking into account the necessary boundary conditions.

In the spirit of the 1-D model, we assume that electron diffusion in the lateral, $y$, direction can be ignored so that electron trajectories are parallel to the $x$ axis (see Figure~\ref{fig:setup}). In this approximation, the effects of bars and dots become identical, as both can be modelled as thickened regions of the bilayer.
We make the following assumptions: in the transverse direction, $z$, we assume that the superconducting film is sufficiently thin that its properties are spatially invariant, even if the film is a proximity bilayer. \cite{Martinis2000} We denote the composite bilayer simply as $\Sp$, $\Spp$ as required, but we include the effect of thickness changes. We ignore pair-breaking by the current. In the lateral, $y$, direction we assume that the film is uniform. The problem is then reduced to the $x-$dimension. In order to present the most general form of our model and its predictions, we have not assumed any particular material parameters for the bias leads, TES or bars. We have used normalized parameters and dimensions wherever possible to emphasize the generic nature of the study.

The ultimate aim of this modelling work is to better inform the detailed design of TESs. By using a model to test possible device designs, we can develop TESs with the required performance parameters, reducing the need for intermediate experimental measurements.

In Section \ref{sec:theory}, we outline the theoretical basis of the model, and describe the numerical method in Section \ref{sec:model}. Section \ref{sec:results} shows a number of outcomes of the modelling and draws qualitative comparisons with existing experimental data. Section \ref{sec:conclusions} summarizes the work.

\section{Theory}
\label{sec:theory}
\begin{figure}
\includegraphics[width=8.6cm]{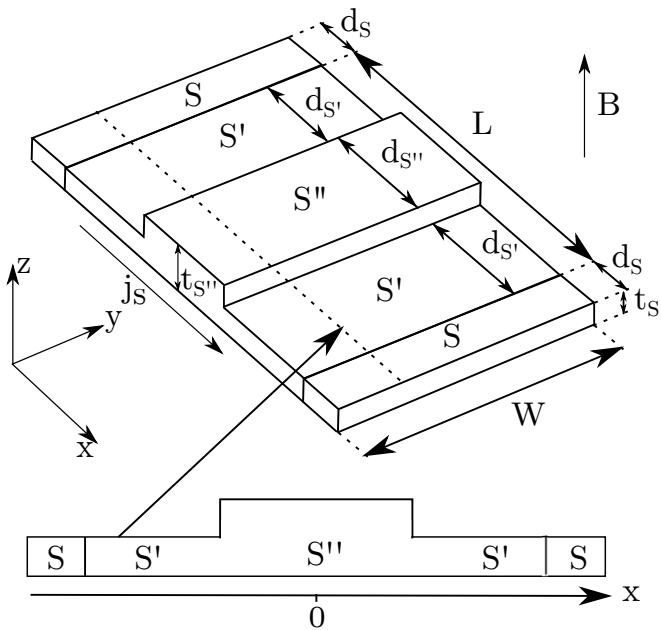}
\caption{\label{fig:setup} The geometry used to model a TES with a single bar. This is given for illustrative purposes - the techniques described are applied to TESs with multiple metal bars. The axes indicate the coordinate system used to describe the sensor and the cut-through shows the one-dimensional model. The regions labelled S are the superconducting electrodes, the bilayer is labelled $\Sp$ and the region with the additional metal bar is labelled $\Spp$. $L$ denotes the length of the sensor, and $W$ its width. $d_i$ indicates the length of region i and $t_i$ its thickness. $B$ shows the direction in which the magnetic field is applied. We assume the supercurrent density $j_{\Sl}$ is in the $x$ direction as indicated.}
\end{figure}

We model the TES in one dimension as shown by the dashed line in Figure \ref{fig:setup}. We follow (2)-(4) from Vasenko et al. \cite{Vasenko2008}, parameterizing the normal and anomalous Green's functions as $G= \cos \theta$ and $F=\sin \theta e^{i\chi}$ respectively, where $\theta$ is the pairing angle and $\chi$ is the superconducting phase. Using matrix notation for the Green's function
\begin{equation}
\label{eq:GF}
\hat{G} = \left(
\begin{array}{l l}
G & F \\
F^* & -G
\end{array} \right)
= \left(
\begin{array}{l l}
\cos \theta & \sin \theta \rme^{\rmi \chi} \\
\sin \theta \rme^{-\rmi \chi} & - \cos \theta
\end{array} \right),
\end{equation}
we obtain the one-dimensional Usadel equations \cite{Vasenko2008,Kozorezov2011,Golubov2004}
\begin{subequations}
\label{eq:finalus}
\begin{align}
\hbar D &\left(\frac{\Diff2 \theta}{\diff x^2} - \left(\frac{\diff \chi}{\diff x} \right)^2 \cos \theta \sin \theta \right) \nonumber \\
&= 2\hbar \omega \sin \theta - \cos \theta (\Delta \rme^{-\rmi \chi} + \Delta^* \rme^{\rmi \chi})  \label{eq:finalus1}, \\
\hbar D & \frac{\diff}{\diff x} \left(\sin^2 \theta \frac{\diff \chi}{\diff x} \right) = \rmi(\Delta \rme^{-\rmi\chi} - \Delta^* \rme^{\rmi\chi}) \sin \theta \label{eq:finalus2},
\end{align}
\end{subequations}
where $\Delta$ is the spatially varying superconducting order parameter. $\omega$ represents the Matsubara frequencies given by $\hbar \omega = \pi k_{\mathrm{B}}  T(2n+1)$ for integer $n$. The Usadel equations \eqref{eq:finalus} make no assumptions about the relationship between the phase of $\Delta$ and the superconducting phase $\chi$.

The order parameter varies as a function of $\chi$ and $\theta$ according to the self-consistency equation
\begin{equation}
\label{eq:final1d}
\Delta \ln \left(\frac{T}{T_{\crit}} \right) + 2 \pi k_{\mathrm{B}} T \sum_{\omega > 0} \left( \frac{\Delta}{\hbar \omega} - \sin \theta \rme^{\rmi \chi} \right) = 0,
\end{equation}
and the supercurrent density is
\begin{equation}
\label{eq:finaljs}
j_{\mathrm{s}}(x,\psi)= -\frac{2\sigma \pi k_{\mathrm{B}} T}{e} \sum_{\omega >0} \sin^2 \theta \frac{\diff \chi}{\diff x}.
\end{equation}
$\psi =\chi(L/2 + d_{\Sl})-\chi(-(L/2 + d_{\Sl}))$ is the phase difference between the superconducting leads.

The Usadel equations \eqref{eq:finalus} and \eqref{eq:final1d} need to be solved with appropriate boundary conditions. We assume bulk superconductor behaviour at the external boundaries $\pm (L/2 + d_{\Sl})$, the outer edges of the superconducting electrodes, so here $\Delta=\Delta_{\text{BCS}}$, $\theta = \atan\left( \Delta_{\text{BCS}}/\omega \right)$ and $\partial \chi/ \partial x =0$. We set
$\chi(\pm (L/2 + d_{\Sl}))=\pm\psi/2$.\cite{Golubov2004,Vasenko2008}
The boundary conditions at the S-$\Sp$ and $\Sp$-$\Spp$ internal interfaces are given in terms of the Green's functions of \eqref{eq:GF} by \cite{Kuprianov1988,Vasenko2008}
\begin{equation}
\label{eq:GBC1}
\xi_{j} \gamma \frac{t_j}{t_i} \left( \hat{G}_{j} \frac{\partial}{\partial x} \hat{G}_{j} \right) = \xi_i \left( \hat{G}_i \frac{\partial}{\partial x} \hat{G}_i \right),
\end{equation}
and
\begin{equation}
\label{eq:GBC2}
2 \xi_{j} \gamma_{\mathrm{B}}  \left( \hat{G}_{j} \frac{\partial}{\partial x} \hat{G}_{j} \right) = \pm[\hat{G}_i, \hat{G}_{j} ].
 \end{equation}
In the direction of increasing $x$, the positive sign in \eqref{eq:GBC2} refers to the interface from material $i$ to material $j$ and the negative sign to the interface from material $j$ to material $i$.  The coherence length is $\xi_i^2 = \hbar D_i/2\pi k_{\mathrm{B}} T_{\crit,i}$, where $D_i$ is the diffusion coefficient. At temperatures above $T_{\crit,i}$, this becomes a correlation length $\xi_i^2 = \hbar D_i/2\pi k_{\mathrm{B}} T$. The material interfaces are characterized by two parameters: $\gamma_{\mathrm{B}}$ is a measure of the boundary resistance;\cite{Kuprianov1988} and $\gamma$ describes the strength of suppression of superconductivity in S near the interface compared to the bulk. \cite{Golubov2004}

We modify the first boundary condition \eqref{eq:GBC1} compared with \onlinecite{Kuprianov1988} with an additional factor $t_{j}/t_{i}$. This takes into account the discontinuities in thickness between the $\Sp$ and $\Spp$ regions, ensuring conservation of supercurrent.

With zero applied magnetic field, the phase difference between the $\Sl$-electrodes, $\psi$, is constant as a function of $y$, and the total supercurrent
\begin{equation}
\label{eq:total_current_no_field}
I_{\mathrm{s}}(x,\psi)=W t (x) j_{\mathrm{s}}(x,\psi)
 \end{equation}
 follows directly by integration with respect to $y$. This spatially varying supercurrent is useful for comparing the characteristics of TES with different geometries. We use the value of $\psi=\psi_0$ that maximizes $I_s$.
In the presence of a magnetic field $B_z$, the gauge-invariant phase difference between the electrodes varies as a function of $y$ such
that\cite{Cuevas2007,Heida1998,Chiodi2012}
\begin{equation}
\label{eq:magphase}
 \psi(y)=\frac{2\pi \Phi(y)}{\Phi_0} + \psi_0,
\end{equation}
where $\Phi_0=h/2e$ is the quantum flux and $\Phi(y)=B_z y (L+2\lambda_{L})$, with $\lambda_{L}$ the 
penetration depth in $\Sl$.

Now the supercurrent density varies as $j_s(x,\psi(y))$, so defining the critical current density as $j_{\crit}(\psi)=\mathrm{min}_x (j_s(x,\psi))$
we find the total critical current
\begin{equation}
\label{eq:supercurrentB}
I_{\crit}(\Phi) = t \int_{y=-L/2}^{y=L/2} {j_{\crit}(\psi(y))} \diff y.
\end{equation}
 It is only necessary to calculate $j_{\crit}(\psi)$ for $\psi \in (0, \pi) $, due to the properties of the current-phase relation.\cite{Golubov2004} This analysis assumes a relatively weak magnetic field, and so the order parameter $\Delta$ does not change.

We model the overall TES resistance using a resistively shunted Josephson junction (RSJ) model. \cite{Kozorezov2011a,Coffey2008} The resistance is given by
\begin{equation}
\label{eq:resistance}
R(T,I) = R_{\nn} \left\{ 1 + \frac{1}{\kappa} \Im \left[ \frac{\mathcal{I}_{1+\rmi \zeta \kappa}(\zeta)}{\mathcal{I}_{\rmi \zeta \kappa}(\zeta)} \right] \right\},
\end{equation}
where $\zeta=\hbar I_{\crit}/2eT$ describes the effect of thermal fluctuations, and $\mathcal{I}_{\mu}(\nu)$ are modified Bessel functions of the first kind of complex order $\mu$ and real variable $\nu$. The ratio of current to critical current is $\kappa=I/I_{\crit,T}$. The Bessel functions are calculated using a continued fraction method.\cite{Gautschi2016}

\section{Model}
\label{sec:model}

We solve the Usadel equations \eqref{eq:finalus} iteratively and check for convergence of the order parameter $\Delta(x=0)$ to better than 0.1$\%$. We assume throughout that $T_{\crit,\Sp}/T_{\crit,\Sl}=0.01$ and $T_{\crit,\Spp}/T_{\crit,\Sl}=0.005$. We take $\gamma_{(\Sl,\Sp)}=0.1$, $\gamma_{\mathrm{B}(\Sl,\Sp)}=1$ to describe the interfaces of S-$\Sp$, and $\gamma_{(\Sp,\Spp)}=1$, $\gamma_{\mathrm{B}(\Sp,\Spp)}=0$ to describe the interfaces of $\Sp$-$\Spp$. Where additional $\Spp$ features are included, we denote their number by $N$ and we assume that they are uniformly positioned  with respect to the centre of the TES bilayer. This method is general and will work for any phase difference $\psi$ between the two electrodes. We use the converged values of $\theta$ and $\chi$ to calculate the supercurrent in material $i$ according to \eqref{eq:finaljs}. From this, we calculate the $R(T,I)$ surface using \eqref{eq:resistance} and hence $\alpha$ and $\beta$.


\section{Results}
\label{sec:results}
\begin{figure}
\includegraphics[width=8.6cm]{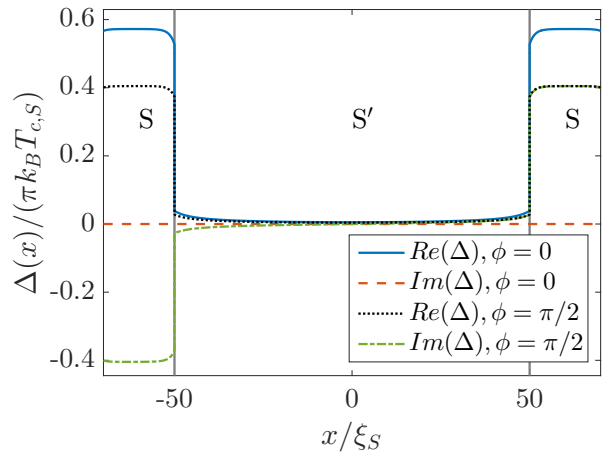}
\caption{\label{fig:deltas} Real and imaginary parts of $\Delta(x)$ across a bare TES with $L/\xi_{\Sl} = 100, T/T_{\crit,\Sl}=0.02.$ The phase difference between the leads is $\psi=0$ for the solid and dashed lines and $\psi=\pi/2$ for the dotted and dot-dashed lines. The units used are dimensionless: $\Delta$ is normalised by a factor of $\pi k_B T_{\crit,\Sl}$  and $x$ by a factor of $\xi_{\Sl}$.}
\end{figure}
\begin{figure}
\includegraphics[width=8.6cm]{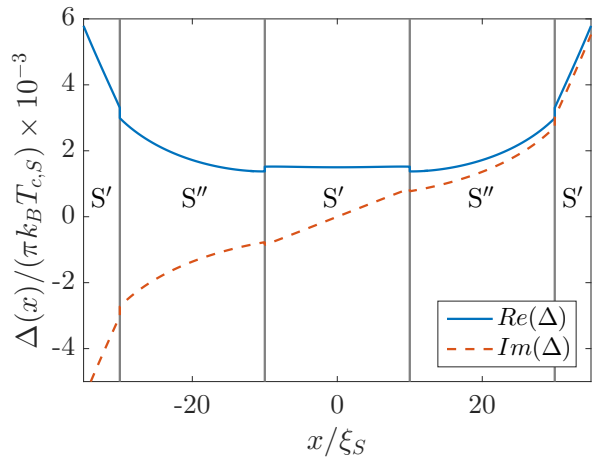}
\caption{\label{fig:deltaNN} Real (solid line) and imaginary (dashed line) parts of $\Delta(x)$ in the central region of a sensor with $N=2$, $\psi=\pi/2$, $T/T_{\crit,\Sl}=0.02$.}
\end{figure}
\begin{figure}
\includegraphics[width=8.6cm]{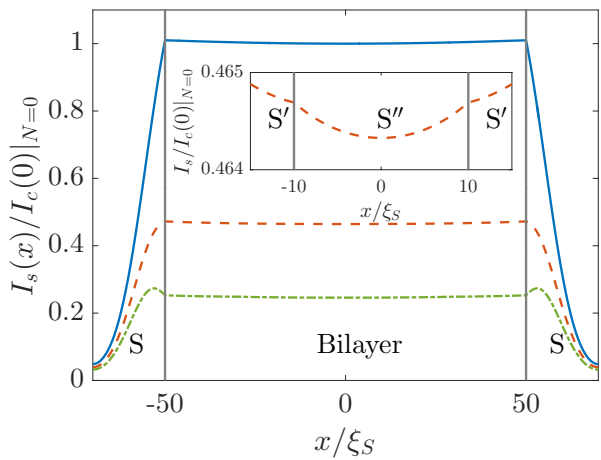}
\caption{\label{fig:jsx} Supercurrent $I_{\mathrm{s}}$ for TESs with $N=0$ (solid line), 1 (dashed line) and 2 (dot-dashed line). Inset is a magnification of the central region of the plot for $N=1$ to illustrate the effects of the lower-$T_{\crit}$ $\Spp$ structure. The phase difference of $\psi=\pi/2$ means that the supercurrent is close to its maximum value; and we take $T/T_{\crit,\Sl}=0.02$. $I_{\mathrm{s}}$ is normalized to $I_{\crit}$ in a sensor with $N=0$ and no magnetic flux ($ I_{\crit}(0)|_{N=0}$)}
\end{figure}
Here we present representative results of the modelling. Figure \ref{fig:deltas} shows the variation of the superconducting order parameter $\Delta$ with position for a TES with $L/\xi_{\Sl}=100, N=0$. The phase differences of 0 and $\pi/2$ were chosen for comparison to the earlier calculations of Kozorezov et al. and show good qualitative agreement with the results in \onlinecite{Kozorezov2011}. Although the temperature used here, $T/T_{\crit,\Sl}=0.02$, is above the superconducting transition temperature of $\Sp$, there is still a non-zero order parameter throughout the length of the sensor due to the long-range proximity effect. Figure \ref{fig:deltaNN} shows the effect on the order parameter of adding two $\Spp$ sections each of length $d_{\Spp}/\xi_{\Sl}=20$. The magnitudes of both the real and imaginary parts of the order parameter are reduced inside these bars.

Figure \ref{fig:jsx} shows how the supercurrent $I_{\mathrm{s}}$ varies with position in a TES of fixed overall length $L/\xi_{\Sl}=100$ with $N=0, 1, 2$ bars of $d_{\Spp}/\xi_{\Sl}=20$. $I_{\mathrm{s}}(x)$ is normalized by the critical current for a bare TES with zero applied field $I{_{\crit}}(0)\vert_{N=0}$. As $N$ is increased, both the overall supercurrent and hence the critical current are reduced. We also see in the inset how the supercurrent magnitude is reduced within the metal bars themselves. However, consistent with our modified boundary conditions at the S-$\Sp$ and $\Sp$-$\Spp$ interfaces \ref{eq:GBC1}, the current $I_{\mathrm{s}}$ is continuous. At the far edges of S, $I_{\mathrm{s}} \rightarrow 0$ as required by the boundary conditions.
 \begin{figure}
\includegraphics[width=8.6cm]{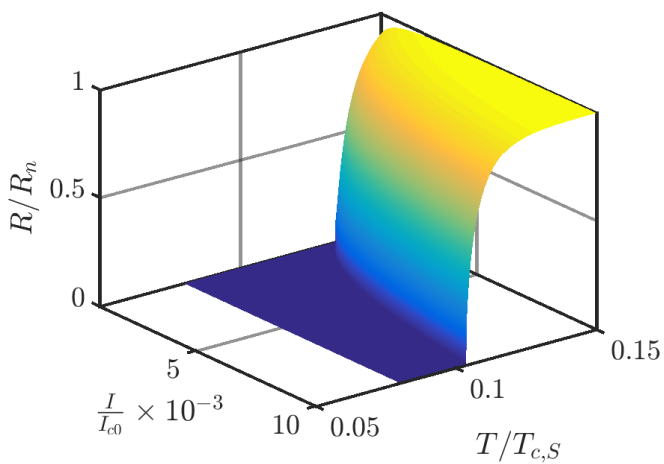}
\caption{\label{fig:RTI} $R(T,I)$ for a TES with $N=0, \psi=\pi/2$. The temperature is normalised to $T_{\crit,\Sl}$, the superconducting transition temperature of S; the current to $I_{\crit 0}$, the critical current in the sensor for $T/T_{\crit,\Sl}=0.05$; and the resistance to $R_{\nn}$, the normalised TES resistance.}
\end{figure}
\begin{figure}
\includegraphics[width=8.6cm]{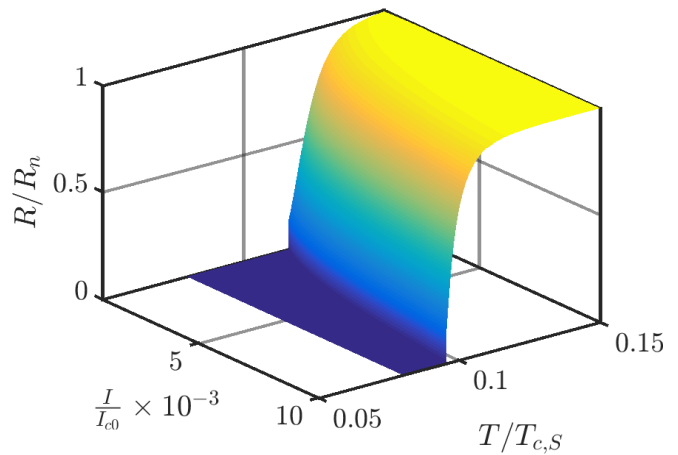}
\caption{\label{fig:RTI2} $R(T,I)$ for a TES with $N=2, \psi=\pi/2$.}
\end{figure}
\begin{figure}
\includegraphics[width=8.6cm]{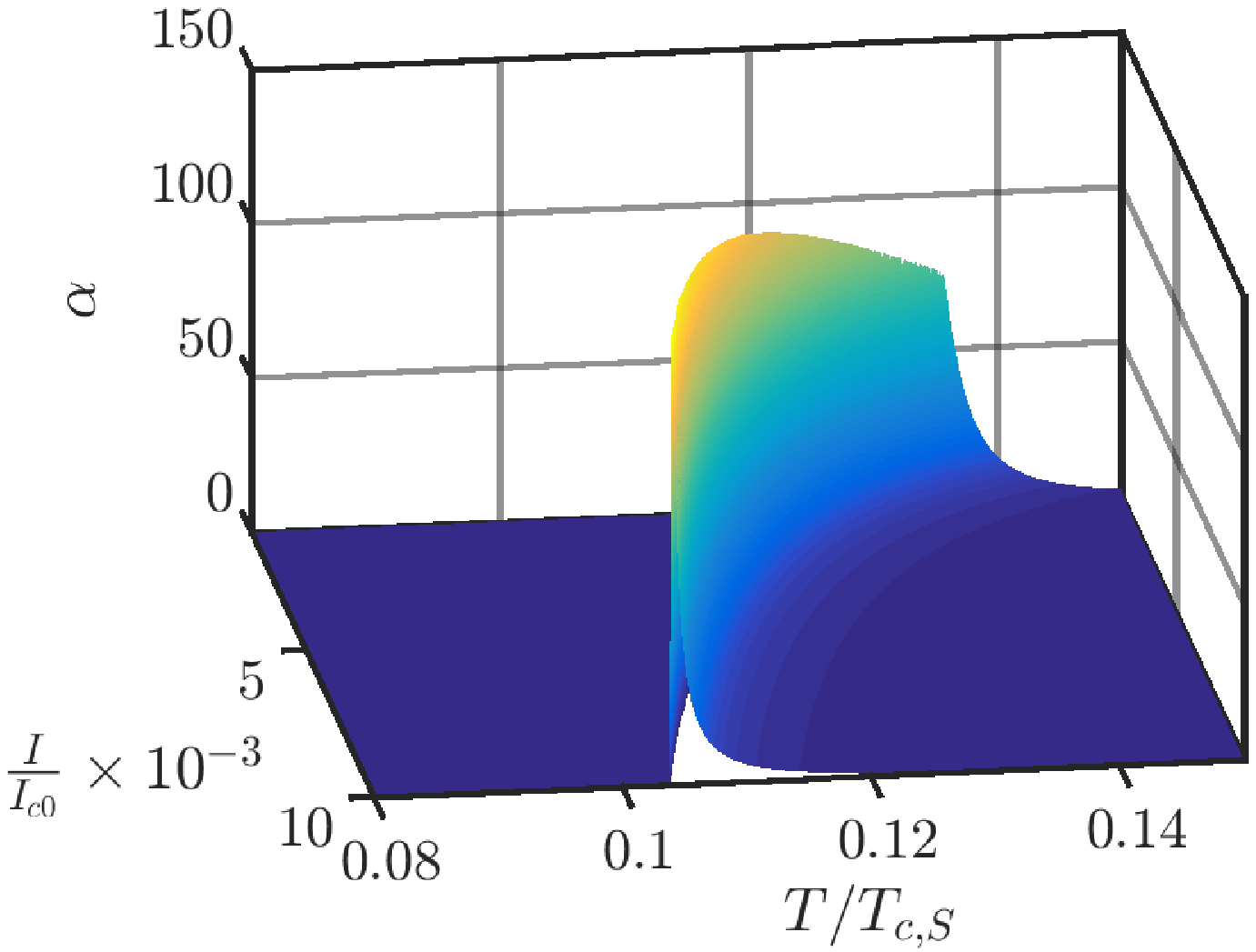}
\caption{\label{fig:alpha0} $\alpha(T,I)$ for a TES with $N=0, \psi=\pi/2$.}
\end{figure}
\begin{figure}
\includegraphics[width=8.6cm]{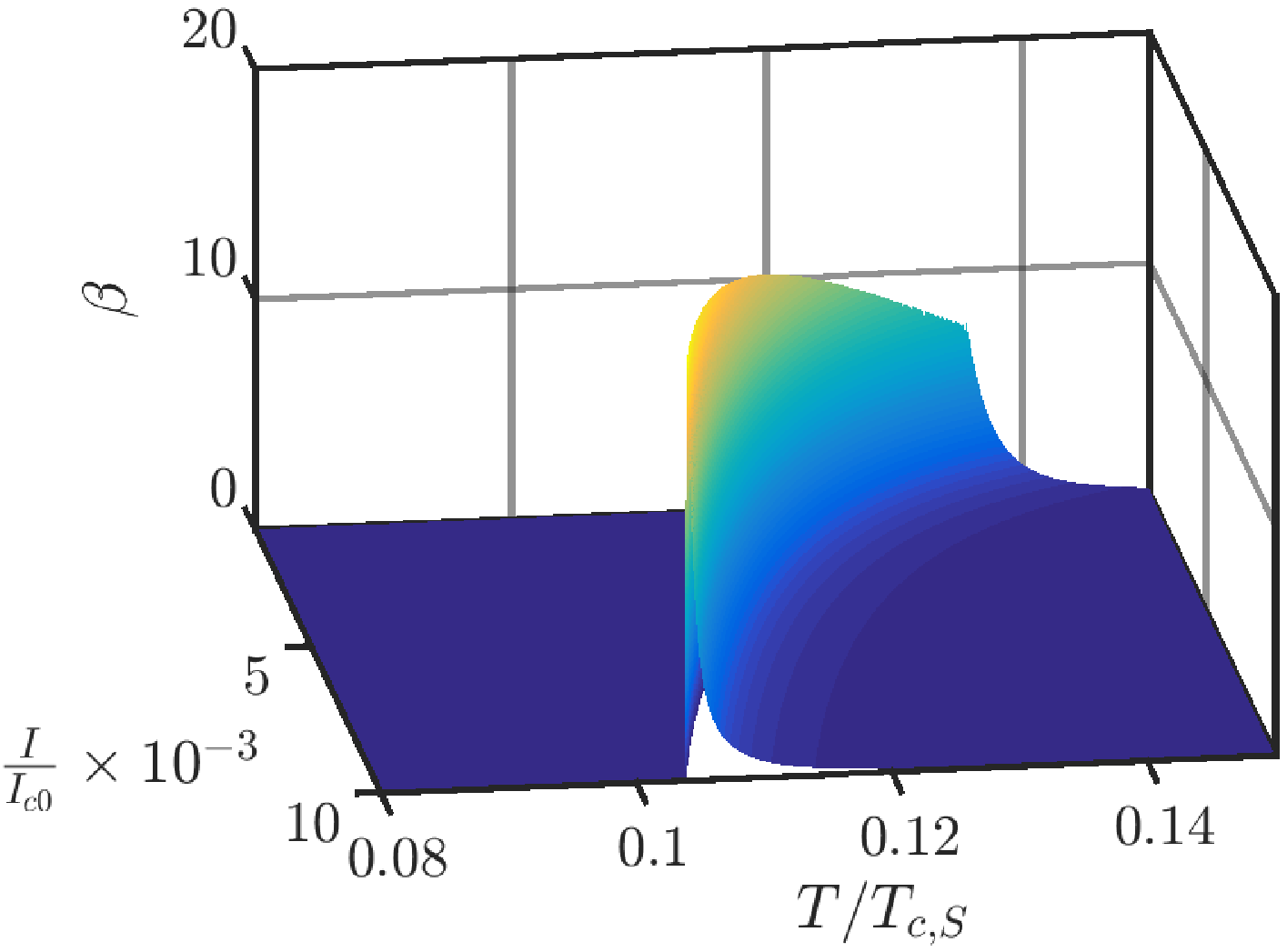}
\caption{\label{fig:beta0} $\beta(T,I)$ for a TES with $N=0, \psi=\pi/2$.}
\end{figure}
\begin{figure}
\includegraphics[width=8.6cm]{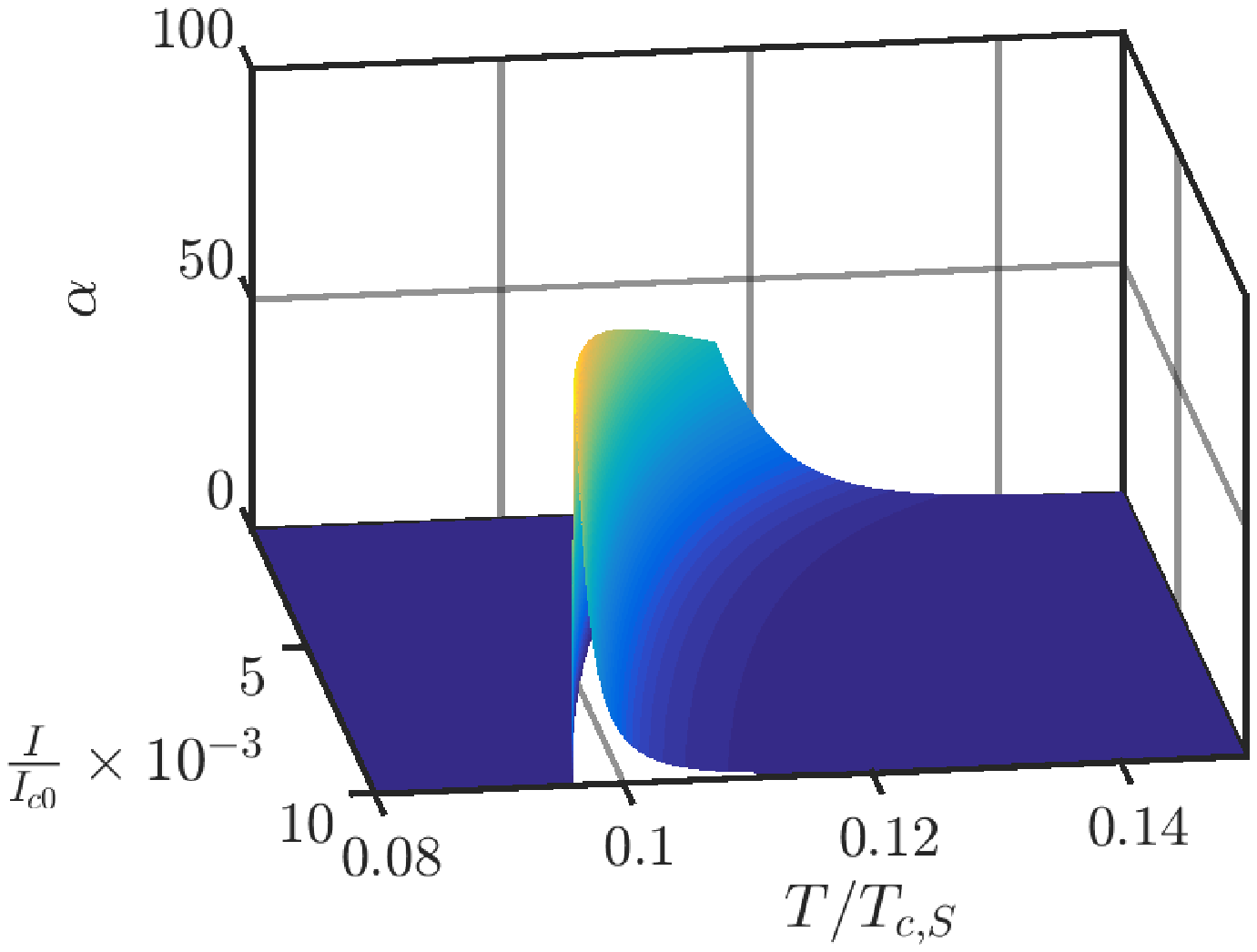}
\caption{\label{fig:alpha2} $\alpha(T,I)$ for a TES with $N=2, \psi=\pi/2$.}
\end{figure}
\begin{figure}
\includegraphics[width=8.6cm]{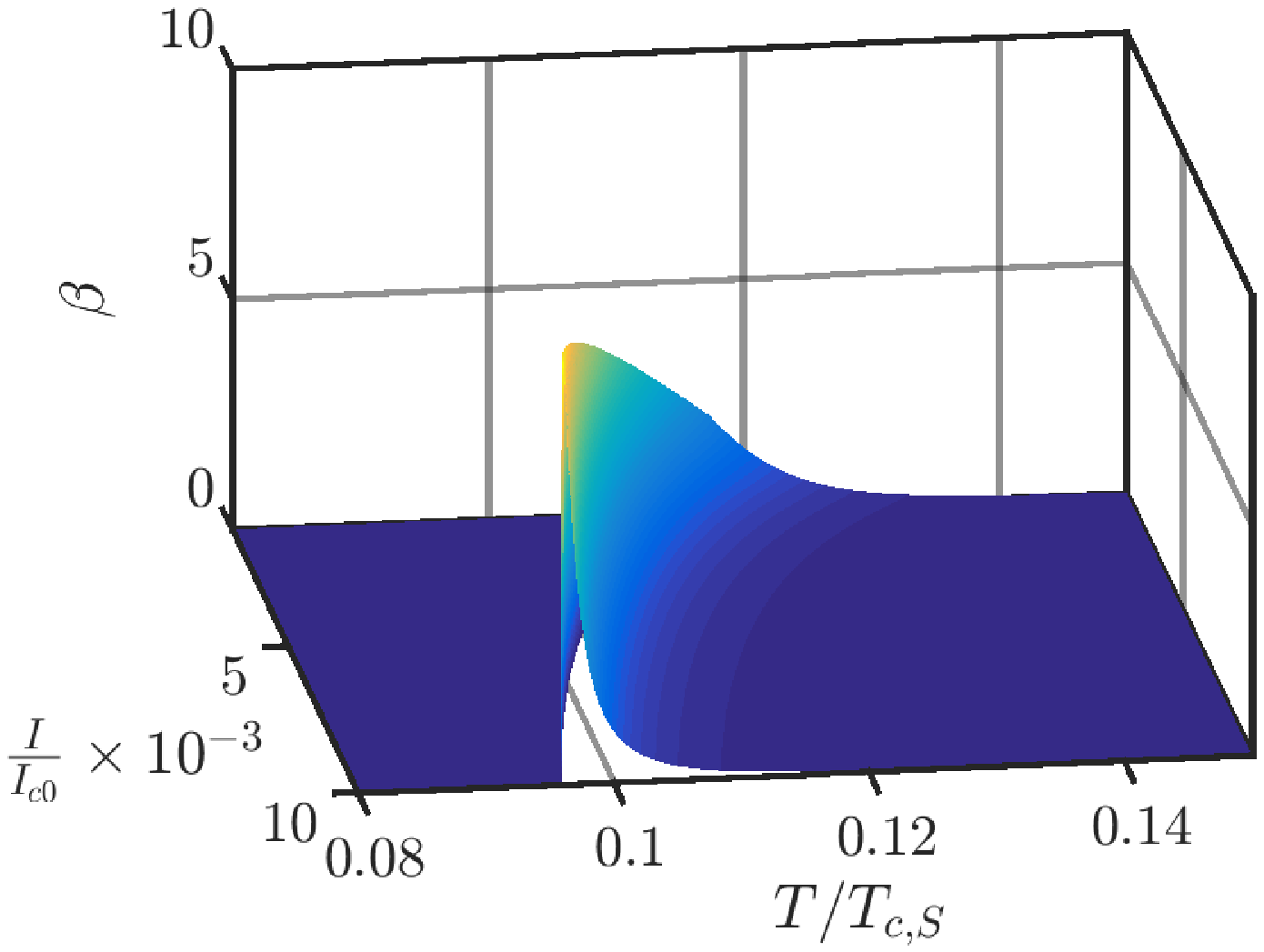}
\caption{\label{fig:beta2} $\beta(T,I)$ for a TES with $N=2, \psi=\pi/2$.}
\end{figure}
\begin{figure}
\includegraphics[width=8.6cm]{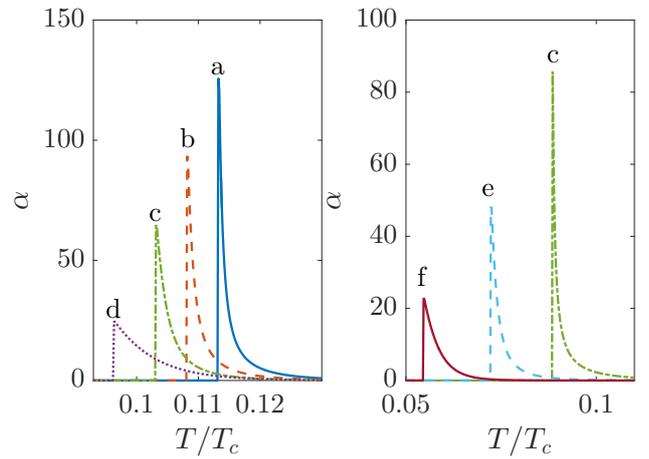}
\caption{\label{fig:alpha_beta} Left panel shows the effects on $\alpha(T)|_{I}$ of increasing the number of bars and the magnetic flux: (a) $N=0, \Phi/\Phi_0=0$; (b) $N=1, \Phi/\Phi_0=0$; (c) $N=2, \Phi/\Phi_0=0$; (d) $N=2, \Phi/\Phi_0=\pi/2$. $L/\xi_{\Sl} =100$ and $I/I_{\crit 0}=0.005$ throughout. Right panel shows the effect on $\alpha(T)|_{I}$ of lengthening the sensor: (c) $L/\xi_{\Sl} =100$ (e) $L/\xi_{\Sl} =150$; and (f) $L/\xi_{\Sl} =200$. $\Phi=0,N=2,I/I_{\crit,0}=0.1$ throughout. $\beta$ shows the same trends as $\alpha$ - from Figures \ref{fig:alpha0} to \ref{fig:beta2}, $\alpha/\beta \approx 10$.}
\end{figure}

Figure \ref{fig:RTI} shows the $R(T,I)$ surface normalised to $R_{\nn}$ as a function of normalised temperature $T/T_{\crit,\Sl}$ and normalised current $I/I_{\crit 0}$. $I_{\crit 0}$ is the critical current in the sensor for $T/T_{\crit,\Sl}=0.05$. The sensor has $N=0, L/\xi_{\Sl}=100$. The effective superconducting transition occurs at about $T/T_{\crit,\Sl}\simeq 0.12$ dependent on current. This is considerably higher than the intrinsic transition temperature of the bilayer, $T_{\crit,\Sp}/T_{\crit,\Sl}=0.01$, due to the proximity effect of the electrodes, the TES geometry and the choice of boundary conditions. Figure \ref{fig:RTI2} shows the calculated $R(T,I)$ surface for a TES with $L/\xi_{\Sl}=100$ and $N=2$. Both lower-$T_{\crit}$ bars have length $d_{\Spp}/\xi_{\Sl}=20$. A reduction in the effective $T_{\crit}$ relative to Figure \ref{fig:RTI} is immediately evident. There is also a suggestion that the $R(T,I)$ surface becomes less steep.

The effect on $\alpha(T,I)$ and $\beta(T,I)$ for the same geometries is shown in Figures \ref{fig:alpha0} to \ref{fig:beta2}. Figure \ref{fig:alpha0} shows the logarithmic resistance-temperature sensitivity $\alpha(T,I)$ with $N=0$ and phase difference $\psi=\pi/2$. Figure \ref{fig:beta0} shows $\beta(T,I)$ for the same parameters. We compare these with Figures \ref{fig:alpha2} and \ref{fig:beta2}, which show the same surfaces for a TES with $N=2$.
A reduction in both $\alpha$ and $\beta$ caused by the bars is evident and we also see a ratio $\alpha / \beta \approx 10 $, agreeing with experimental observations.\cite{Lindeman2006,Goldie2009,Smith2013}

These trends are more fully shown in Figure \ref{fig:alpha_beta}. The left panel shows the effects on $\alpha(T)|_{I}$ of increasing the number of bars and the magnetic flux. In (a) $N=0, \Phi/\Phi_0=0$; (b) $N=1, \Phi/\Phi_0=0$; (c) $N=2, \Phi/\Phi_0=0$; (d) $N=2, \Phi/\Phi_0=\pi/2$. $L/\xi_{\Sl} =100$ and $I/I_{\crit,0}=0.005$ throughout. The right panel shows the effect on $\alpha(T)|_{I}$ of lengthening the sensor: (c) $L/\xi_{\Sl} =100$ (e) $L/\xi_{\Sl} =150$; and (f) $L/\xi_{\Sl} =200$. $\Phi=0,N=2,I/I_{\crit,0}=0.1$ throughout. Comparison of curves (a) to (c) shows how, as observed experimentally\cite{Wang2015,Ullom2015}, $\alpha$ and hence $\beta$ are reduced as $N$ increases. Comparison of curves (c) and (d) shows the reduction of $\alpha$ and hence $\beta$ by an applied field. Comparison of curves (c), (e) and (f) shows the effect of increasing the TES length with fixed $N$. The effective $T_{\crit}$ is reduced by increasing either $N$ or $L$ and we find that $T_{\crit}\propto L^{-0.7}$ in this instance with $N=2$. These predictions enable us to quantify the effects of adding bars and lengthening the sensor on key device parameters.

\begin{figure}
\includegraphics[width=8.6cm]{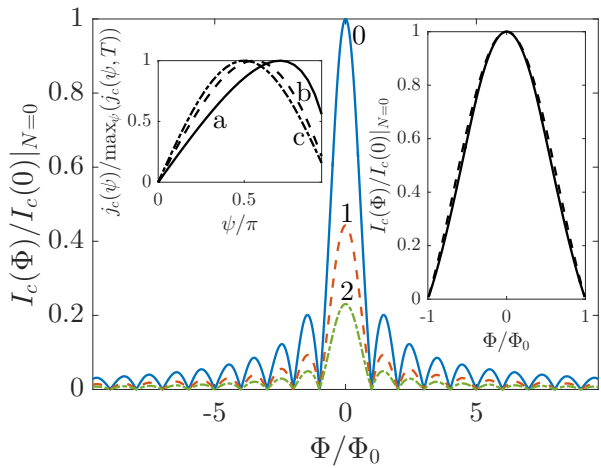}
\caption{\label{fig:ISB} Critical current in the sensor $I_{\crit}$ as a function of magnetic flux $\Phi$, for TESs with (0) $N=0$, (1) $N=1$ and (2) $N=2$. Magnetic field is applied perpendicular to the sensor, in the $z$ direction, $T/T_{\crit,\Sl}=0.02$ and $L/\xi_{\Sl}=100$.  $I_{\crit}$ is normalized to the zero-field critical current for a bare TES, $I_{\crit}(0)|_{N=0}$, and $\Phi$ is in units of the magnetic flux quantum $\Phi_0$. Left inset shows $j_{\crit}/\mathrm{max}_{\psi}(j_{\crit}(\psi,T))$, the critical current normalized to its maximum value, as a function of phase difference $\psi$ across the sensor, at $T/T_{\crit,\Sl}=$ (a) 0.01, (b) 0.02 and (c) 0.05. Right inset compares the normalised critical current for $N=0$ (solid line) with a sinc function of the same period and magnitude (dashed line).}
\end{figure}

Figure \ref{fig:ISB} shows the variation of critical current with magnetic flux $I_{\crit}(\Phi)$, normalized to $I_{\crit}(0)|_{N=0}$. The predicted critical current-phase relationships have the form of a Fraunhofer diffraction pattern even with the addition of lower-$T_{\crit}$ features. The calculated current-phase relationship  $j_{\crit}(\psi)$ is not sinusoidal, showing the expected temperature-dependent skew.\cite{Likharev1979, Golubov2004} The maximum of $j_{\crit}(\psi)$ increases with phase as the temperature is reduced, as shown in the left inset to Figure \ref{fig:ISB}. As described in Section \ref{sec:theory}, we use this $j_{\crit}(\psi)$ relationship calculated directly from the Usadel equations to determine the field dependence. The right inset demonstrates that the sinc form is not exact showing deviations of order $5\%$ from a sinc function of the same period and magnitude. Such Fraunhofer-like flux-dependencies have been observed experimentally. \cite{Sadleir2011,Gottardi2014a} 

\begin{figure}
\includegraphics[width=8.6cm]{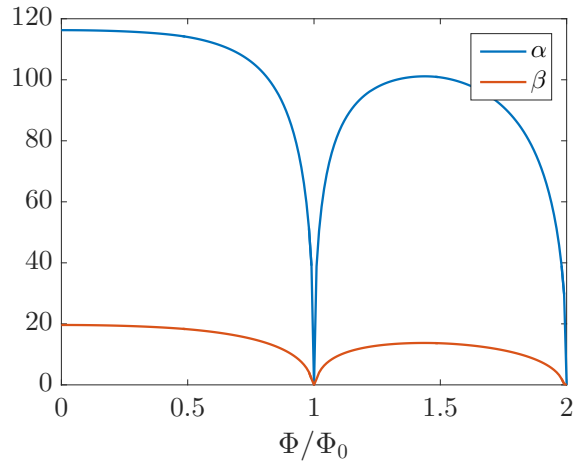}
\caption{\label{fig:alpha_beta_Phi} $\alpha$ and $\beta$ as a function of magnetic flux in units of the magnetic flux quantum $\Phi/\Phi_0$. The calculations of both parameters are carried out for a resistance of $R/R_{\nn}=0.015$. For $\alpha$, $I/I_{\crit 0}=0.8$ and for $\beta$, $T/T_{\crit,\Sl}=0.063$.}
\end{figure}

We also see a periodic dependence of the electrothermal parameters $\alpha$ and $\beta$ on applied magnetic field, as demonstrated by Figure \ref{fig:alpha_beta_Phi}. Again, the relationship is not perfectly symmetric - there is a slight skew in the curve. This behaviour agrees qualitatively with measurements made by Smith et al.\cite{Smith2013} Calculations of this kind will be important in quantifying the effects of stray magnetic fields on device performance.

\section{Discussion and Conclusions}
\label{sec:conclusions}

We have developed a 1-D model of TESs operating at low temperatures $T/T_{\crit,\Sl}\ll 1$ to include the effects of normal-metal structures on the bilayer, describing the sensor in terms of the diffusive Usadel equations. We have presented boundary conditions that account for thickness discontinuities in the TES, ensuring supercurrent conservation. We explore the effects of these bars on the principal TES characteristics: $\alpha$, $\beta$, $T_{\crit}$ and $I_{\crit}(\Phi)$.

We find qualitative agreement with existing experimental measurements for the dependence of $\alpha$, $\beta$ and $T_{\crit}$ on the number of bars for fixed TES length. Our calculated  $\alpha/\beta$ ratios are also in agreement with the measurements. Likewise calculated absolute values of $\alpha$ are very similar to published values. We find a good account of the effect of TES length on $T_{\crit}$ and find a dependence  $T_{\crit}\propto L^{-0.7}$ for the TES parameters used here. We also show that $\alpha$ and $\beta$ display oscillatory behaviour when a magnetic field is applied.

These observations are in general agreement with trends seen in the literature, which is a qualitative validation of the predictive power of our model, and supports the idea that models of this kind can be important in device design. We envisage that our model could be used to validate designs prior to device processing work being undertaken. For example, the ability to check whether the electrothermal parameters and transition temperature meet instrument specifications would be immensely beneficial.

We conclude that our model gives a good first-order account of the behaviour of structured TESs. In view of our success to date, we will investigate implementing a full 2-D model \cite{Amundsen2015} within the context of the Usadel equations.

\bibliography{library}

\end{document}